\documentstyle[a4p,12pt,here,epsf,hangcaption,palatino]{article}
\begin{document}
\begin{flushright}
\begin{minipage}{3.5cm}
KOBE-HEP-96-01\\
May 1996
\end{minipage}
\end{flushright}
\vspace*{2.5cm}
\begin{center}
{\LARGE Measurement of $Br(H\to c\bar c, gg)/Br(H\to b\bar b)$ \\
in ${\rm e}^+{\rm e}^-$ colliders at center-of-mass energy of 300 GeV }\\
\vspace*{1cm}
{\Large Isamu Nakamura} \\ \vspace*{5mm}
{\it Guraduate School of Science and Technology, Kobe University,\\
1-1 Rokko-dai, Nada, Kobe, Hyogo, 657, Japan } \\ \vspace*{5mm}
and \\\vspace*{5mm}
{\Large Kiyotomo Kawagoe} \\ \vspace*{5mm}
{\it Department of physics, Faculty of Science, Kobe University,\\ 
1-1 Rokko-dai, Nada, Kobe, Hyogo, 657, Japan }
\end{center}
\vspace*{2cm}
\begin{abstract}
Once a light Higgs boson is discovered
at a future ${\rm e}^+{\rm e}^-$ collider, 
the next target at the collider will be 
precise measurements of the Higgs boson properties. 
In this paper we report a simulation study on the measurement of 
the ratio $Br(H\to c\bar c, gg)/Br(H\to b\bar b)$ at center of mass
energy of 300 GeV,  
and show the possibility to constrain MSSM parameters
from the measurement.
\end{abstract}
\vspace{2cm}
\begin{center}
{\it ( To appear in Physical Review D)}
\end{center}
\pagebreak
%
%
\section{Introduction}
Next generation ${\rm e}^+{\rm e}^-$ linear colliders, firstly to be
operated at center of mass energies ($\sqrt{s}$\thinspace) of 300-500
GeV (phase 1) and to be then upgraded to 1 TeV or higher (phase 2),
have been proposed by several groups in recent
years~\cite{JLC-I,LC-1,LC-2,LC-3,LC-4}. At the first phase of such
colliders the main target is the Higgs boson, the only undiscovered
participant in the Standard Model (SM). According to the recent
precision electroweak measurements at LEP and SLC, Supersymmetry
(SUSY) has become more favorable.  In the Minimal Supersymmetric
extension of the Standard Model (MSSM), two Higgs doublets are
necessary to give the masses of up and down type fermions. Two neutral
scalars ($h^0,H^0$), a pseudoscalar ($A^0$) and a pair of charged
Higgs ($H^\pm$) are introduced in this extended Higgs sector. The mass
of the lightest neutral Higgs ($m_{h^0}$) is smaller than that of the
$Z^0$ boson ($m_{Z^0}$) at tree level. Even if radiative corrections
are taken into account, $m_{h^0}$ is at most $\sim$ 130 GeV for a top
mass ($m_{t}$) of 170 GeV~\cite{HIGGS-1,HIGGS-2,HIGGS-3}. This
situation does not change significantly in extended SUSY
models~\cite{higgscros}.  Therefore if SUSY is correct, discovery of
at least one neutral Higgs boson is guaranteed at the first phase.

\begin{figure}[H]
\epsfxsize=.55\linewidth
\centerline{\epsfbox{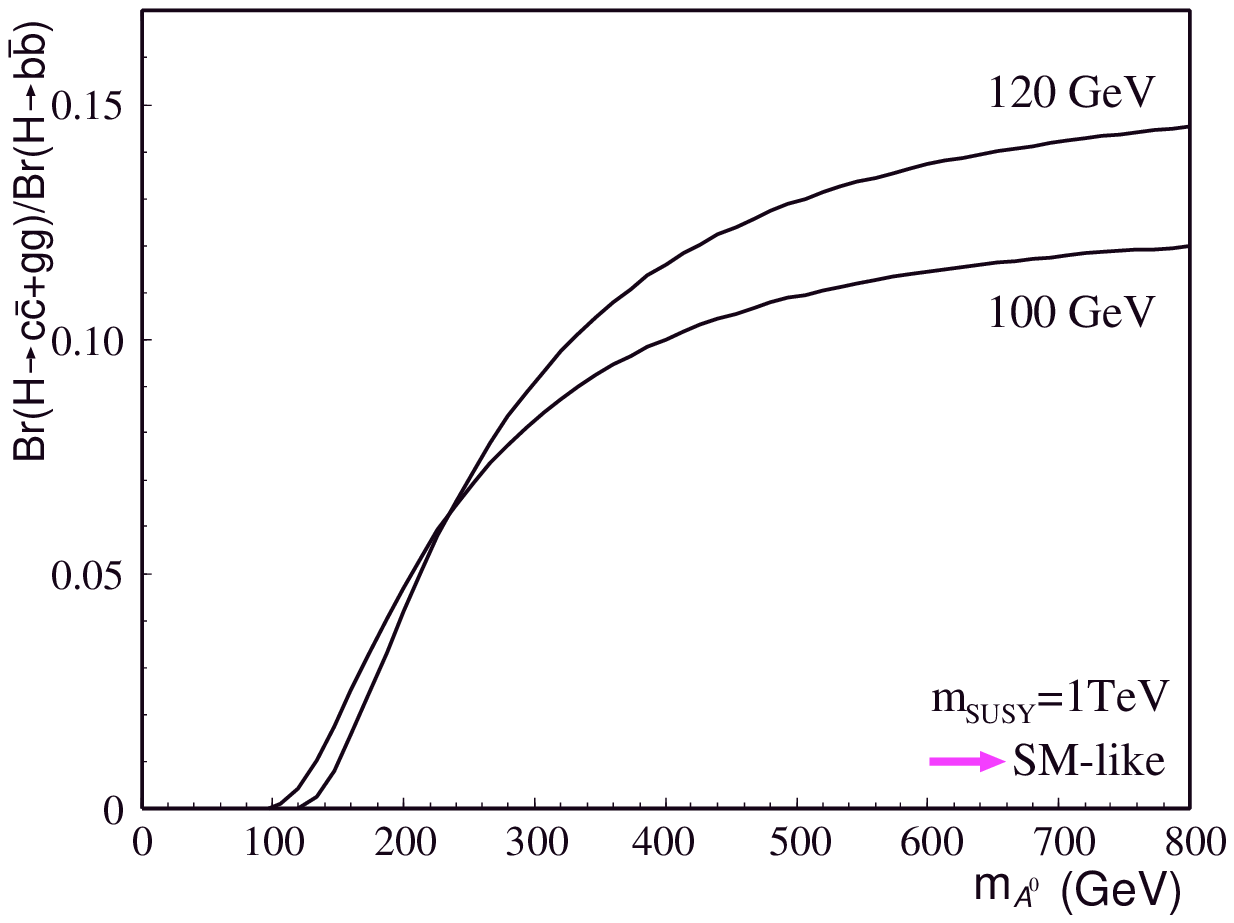}}
\hangcaption{Relation between 
	$Br(H\to c\bar c, gg)/Br(H\to b\bar b)$ and $m_{A^0}$
	~\protect\cite{kamoshita}.}
\label{branch-azero}
\end{figure}

After having discovered the Higgs boson\footnote{%
Hereafter we denote $H$ as the discovered Higgs boson.}, 
the next step is to distinguish
whether the Higgs is SM-like or SUSY-like.  This can be done by
measuring the ratio of branching fractions $Br(H\to c\bar c)/Br(H\to
b\bar b)$ or almost equivalently $Br(H\to c\bar c, gg)/Br(H\to b\bar
b)$\footnote{%
 We define $Br(H\to c\bar c, gg)$ as $Br(H\to c\bar
c)+Br(H\to gg)$.}.
 In the MSSM these values are smaller than those of
the SM.  As the value $Br(H\to c\bar c)/Br(H\to b\bar b)$ (or $Br(H\to
c\bar c, gg)/Br(H\to b\bar b)$) has a strong correlation to the $A^0$
mass ($m_{A^0}$) and almost independent from the SUSY energy scale
($m_{\rm SUSY}$), it is pointed out that a constraint on $m_{A^0}$ can
be obtained from this measurement~\cite{kamoshita}.
Fig.~\ref{branch-azero} shows the relation between $Br(H\to c\bar
c, gg)/Br(H\to b\bar b)$ and $m_{A^0}$.

   The mass of the bottom quark ($m_b$) is important in this relation,
since the ratio $Br(H\to c\bar c, gg)/Br(H\to b\bar b)$ is proportional
to $m_b^{-2}$.  We can use the measurement of
$Br(H\to{\tau}^+{\tau}^-)$ to determine $m_b$, through the relation
$Br(H\to{\tau}^+{\tau}^-)/Br(H\to b\bar b)\propto m_{\tau}^2/m_b^2$
which is valid both in the SM and SUSY models.

   The total production cross section of the process $e^+e^-\to H Z^0$
($\sigma_{tot}$) can be determined by counting the events, in which
the $Z^0$ decays into a pair of leptons whose recoil mass is consistent with
the Higgs mass. This technique is valid even if $H$ has decay channels
to undetectable particles such as neutralinos, since only the $Z^0$ decay
products are used in the event selection.

   A previous simulation study for the measurement of the Higgs boson
branching fractions can be found in Ref.~\cite{hildreth}, 
where the measurement was studied
at $\sqrt{s}=$ 400-500 GeV mainly for $m_{H}=140$ GeV. 
The authors of Ref.~\cite{hildreth}
claimed that the statistical error of the 
$Br(H\to c\bar c, gg)/Br(H\to b\bar b)$ 
measurement would be about 40 \% for the 120 GeV Higgs with
an integrated luminosity of 50 fb$^{-1}$. However, their result is
not sufficient to provide strong constraints on $m_{A^0}$, and
the Higgs mass of 140 GeV may be too heavy if we assume the MSSM.
Therefore we perform here an extended simulation study focusing
on the measurement of the ratio $Br(H\to c\bar c, gg)/Br(H\to b\bar b)$
assuming a SM-like Higgs, where we try to improve the measurement as
follows;
\begin{itemize}
  \item setting $\sqrt{s}=$300 GeV to avoid large background from top
        pair productions 
  \item using all the $Z^0$ decay modes to
        reduce statistical errors 
  \item studying the effect of beam
        polarization to reduce electroweak background.
\end{itemize}
%
%
\section{Event Generation and Detector Simulation}

We use {\sc Pythia5.7 + Jetset7.4}~\cite{pythia} to
calculate the cross sections and to generate events 
of the processes $e^+e^-\to Z^0H$ and major background
processes. 
We use the quark masses of $m_t =$ 170 GeV, 
$m_b =$ 4.25 GeV and $m_c =$
1.35 GeV and initial state radiations are considered.    
A cross check using GRACE~1.1~\cite{grace} shows no
inconsistency in cross sections and angular distributions. 
Table~\ref{cross} shows the list of branching ratios,
cross sections and expected number of events in a year 
(50 fb$^{-1}$) for the SM Higgs boson.
Number of generated events both for signal and background processes
in this study are also listed.
The cross sections with 90 \% polarized electron
beam\footnote{%
Polarization of electron is defined as: 
$P_{\rm e^-} = (N_{\rm L}-N_{\rm R})/(N_{\rm L}+N_{\rm R})$, 
where $N_{\rm L}$ and $N_{\rm R}$ are the number of 
left-handed and right-handed electrons, respectively.}
are obtained from the cross sections and
angular distributions calculated with GRACE~1.1.

The 4-momenta of all decay products given by the event generator 
are fed into a fast detector simulation program 
which is based on a smearing method with the standard parameters 
of the JLC-I Detector given in Ref.~\cite{JLC-I}.
The characteristics of the detector simulator are the following:
\begin{itemize}
\item A pixel type vertex detector (VTX) with a pixel size of 
	$25\times 25 \mu m^2$ is assumed.
\item The momentum resolution of the central drift chamber (CDC)
	is assumed to be $\sigma_{p_t}/p_t = 1.1 \times 10^{-4}
	\cdot p_t \oplus 0.1 \%$ ($p_t$ in GeV).
\item The electromagnetic and hadronic calorimeter hits 
	are created according to the energy resolutions of 
	$15 \%/\sqrt{E} \oplus 1 \%$ and $40 \%/\sqrt{E} \oplus 2 \%$,
	respectively ($E$ in GeV).
\item The momentum measured with the CDC is always used for a charged
	track rather than the energy measured with the calorimeter.
\item The contribution of charged tracks is subtracted
	from the energy of the calorimeter clusters
	in order to extract the energy of neutral particles.
\item All electrons(muons) with momentum above 1 GeV(2 GeV) 
	are assumed to be perfectly identified.
\end{itemize}
\begin{table}[H]
\caption{Branching ratios,
	cross sections, expected numbers of events, expected numbers of
	events with 90 \% polarized electron beam and  numbers of
	generated signal and background events in a year for the SM
	Higgs boson with a mass of 120 GeV (50 fb$^{-1}$). 
	Cross sections for $e^+e^-\to e\nu W,e^+e^-Z^0$ with 90 \%
	polarized electron beam are not calculated here 
	since no events survived the
	selection criteria described in section
	\protect\ref{sec:seleevent}.}   
\label{cross}
\begin{center}
\begin{tabular}{l*{5}{r}} \hline \hline
 $m_{H}$=120 GeV & Br (\%) &  $\sigma$(fb)$\times$Br & events/yr &
 90\% Pol.  &Created\\ \hline
$e^+e^- \to Z^0H$   & ---  & 191   & 9,540   & 7,920  & ---   \\ 
~~~~$H\to b\bar b$    & 66.2& 126 & 6,310   & 5,240  & 50,000 \\ 
~~~~$H\to c\bar c$    & 2.40 & 4.58   & 429     & 190    & 10,000 \\ 
~~~~$H\to gg$         & 8.33 & 15.9  & 794     & 659    & 10,000 \\ 
~~~~$H\to c\bar c+gg$ & 10.7 & 20.5 & 1020    & 850    & 10,000 \\ 
~~~~$H\to\tau^+\tau^-$ & 7.28 & 13.9 & 694     & 576    & 10,000 \\ 
~~~~$H\to W^+W^-$    & 13.4 &  25.6 & 1,280   & 1060   & 10,000 \\ 
\hline
$e^+e^-\to W^+W^-$  & --- & 13,900& 697,000 & 80,200 &700,000 \\ 
$e^+e^-\to Z^0Z^0$   & --- & 1,080 & 53,900  & 36,100 &108,000 \\
$e^+e^-\to e^+e^-Z^0$& --- & 8,310 & 416,000 & ---    &100,000 \\ 
$e^+e^-\to e\nu W$  & --- & 2,290 & 115,000 & ---    &100,000 \\ \hline\hline
\end{tabular}
\end{center}
\end{table}
%
%
\section{Higgs Event Selection}\label{sec:seleevent}
The final states of the process $e^+e^-\to Z^0H$
can be classified into three types in terms of the  $Z^0$ decay modes.
\begin{tabbing}
xxxxx \= xxx \= xxxxxxxx \= xxxxxxxxxxxx \= xxx \kill
\> 1) \> 4 jet \> $Z^0 \to q\bar q$ \> $\sim$ 70 \%\\
\> 2) \> 2 jet \> $Z^0 \to \nu\bar \nu$ \> $\sim$ 20 \%\\
\> 3) \> 2 jet+{\it ll} \> $Z^0\to{\it l}^+{\it l}^-$ \> $\sim$ 10 \%
\end{tabbing}
The event selection for each final state is described 
in this section. 
The selection criteria for the $\sigma_{tot}$ 
determination is described in section~\ref{sec:2ljet}.
All cut values described here are for the Higgs mass of 120 GeV.
Some of the cut values are changed for other Higgs masses 
to optimize the signal/background ratio.
%
%

\subsection{4 Jet Analysis}\label{sec:4jet}

\begin{figure}
\epsfxsize=.55\linewidth
\centerline{\epsfbox{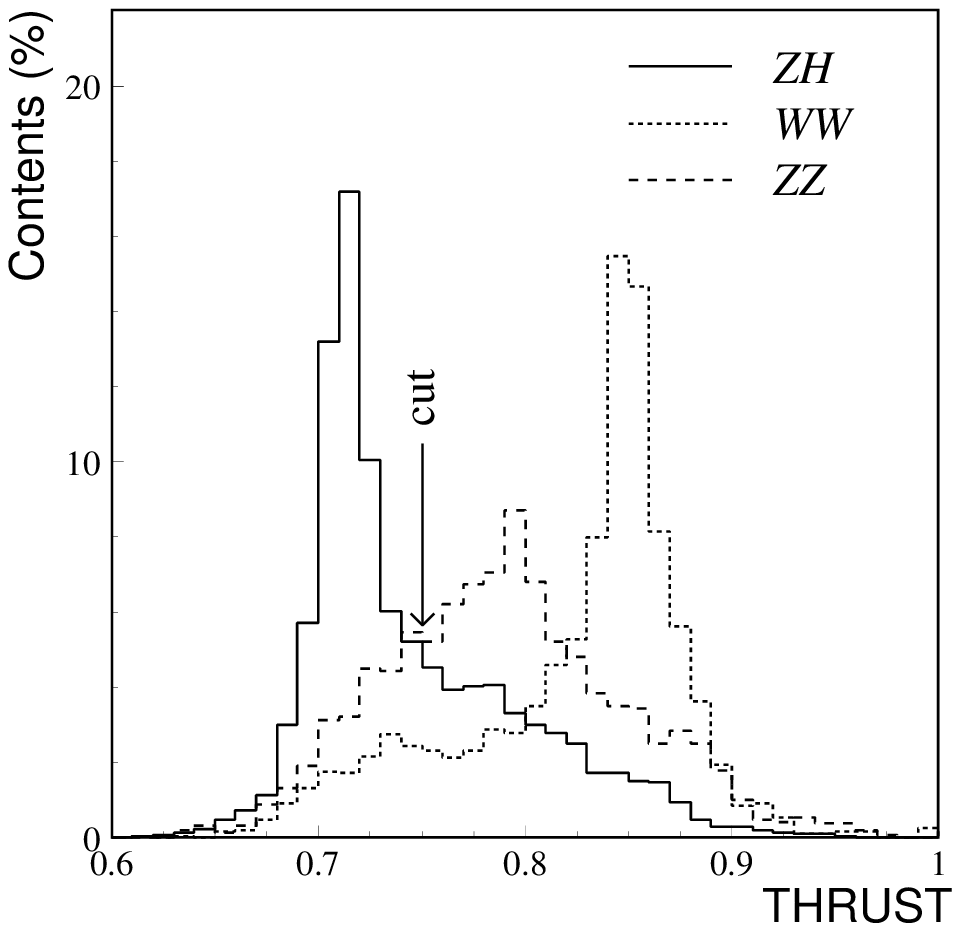}}
\hangcaption{Distribution of $Thrust$ value for the signal and background
	events in 4 jet analysis.}
\label{4jetZH}
\end{figure}

In the 4 jet analysis, each event is forced to be reconstructed as a 4
jet final state (two jets from Higgs and other two from $Z^0$) with
the JADE jet finder~\cite{yclus}.  All possible combinations of jets
are examined to select the combination which minimize the value
\[ (M_{ij}-m_{H})^2+(M_{kl}-m_{Z^0})^2~~~~i,j,k,l=1,2,3,4 ,\]
where $M_{ij}$ is the invariant mass reconstructed 
from the jets $i,j$ assigned to the Higgs jets,
and $M_{kl}$ is that from the jets $k,l$ assigned to the $Z^0$ jets. 
We denote $M_{\rm jet}^H$ and $M_{\rm jet}^Z$ as the reconstructed
Higgs and $Z^0$ masses in the selected combination.
Cuts on the visible energy of $E_{vis} \geq 0.85\cdot\sqrt{s}$ and
on the longitudinal and transverse momentum balances
of $| \sum p_z | \leq$ 30 GeV and $| \sum p_t | \leq$ 30 GeV
are then applied. A containment cut of 
$|\cos{\theta_{\rm thrust}}| \leq 0.8$ and a cut on the thrust value  
$Thrust \leq$ 0.75 are applied to reduce 
$e^+e^-\to W^+W^-,Z^0Z^0$ backgrounds (see Fig.~\ref{4jetZH}). 
Events in which the value $\Delta_{ZH}$ defined as
\[\Delta_{ZH}= 
\left(\frac{M_{\rm jet}^{H}-m_{H}}
{\sigma_{M_{\rm jet}^{H}}}\right)^2 
+ \left(\frac{M_{\rm jet}^{Z}-m_{Z^0}}
{\sigma_{M_{\rm jet}^{Z}}}\right)^2 \] 
is less than 5 are selected as $Z^0H$ candidates.
Once events are selected as $Z^0H$ events,
we apply further cuts to select signal events ($H\to b\bar b,
c\bar c~{\rm and}~gg$).
To reduce $H\to\tau^+\tau^-$ background events,
the number of charged tracks from the Higgs jets 
is required to be greater than 20. This rather large cut value
is chosen because there is inevitable contamination of tracks
from the $Z^0$ jets to the $H$ jets in the jet finding procedure. 
The value $Y_{cut}^{6jet}$, returned from the JADE jet finder for 6
jet final state, must be smaller than 0.005 to reduce 
the $H\to W^+W^-$ background. This cut is based on the fact
that the background event $H\to W^+W^-$ has 6 jets in the final state
while the signal events have only 4 jets.
The expected numbers of events to be collected in a year 
and the selection efficiencies\footnote{%
Selection efficiency is defined as:
$\varepsilon_{\rm sel} = 
\frac{\displaystyle \rm\#~of~selected~events}
{\displaystyle \rm\#~of~generated~events}$}
are summarized in Table~\ref{seleZH}.

%
%
\subsection{2 Jet Analysis}\label{sec:2jet}

\begin{figure}[H]
\epsfxsize=.48\linewidth
\centerline{\epsfbox{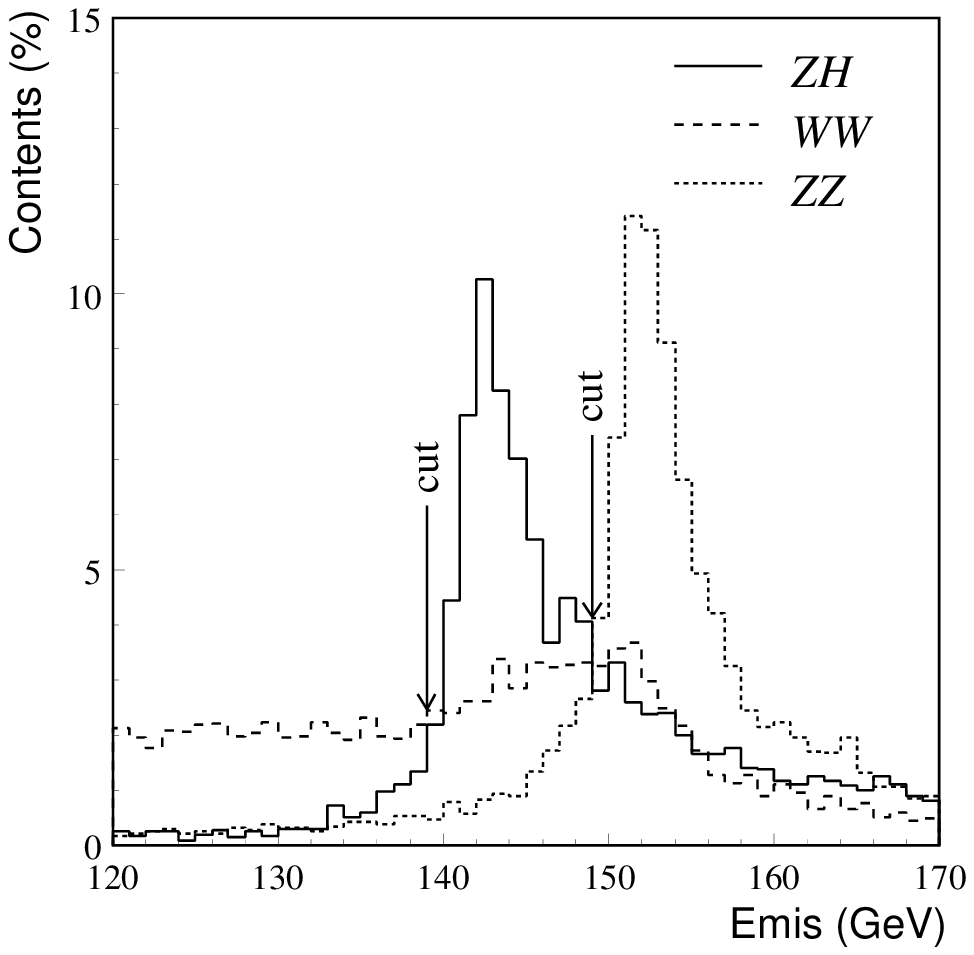}}
\hangcaption{Missing energy distribution for the signal and background
	events for 2 jet analysis.}
\label{2jetZH}
\end{figure}

   In the case where the $Z^0$ decays to $\nu\bar \nu$, the final
state has two jets from the Higgs accompanied by a large missing
energy. Since the reaction $e^+e^-\to Z^0H$ is essentially a two body
process, the missing energy distribution has a sharp peak although
initial state radiations degrade this character.  The events are
required to have a missing energy of 139 GeV $\leq E_{\rm mis} \leq$
149 GeV (see Fig.\ref{2jetZH}).
A containment cut of $|\cos{\theta_{\rm jet}}| \leq $0.8 is also
applied.  Events with high energy leptons ($E_{\rm lepton} \geq$ 10
GeV) are removed. The event is accepted as a $Z^0H$ candidate if the
value $\Delta_{ZH}$ defined as
\[\Delta_{ZH} = 
  \left(\frac{M_{\rm jet}-m_{H}}{\sigma_{M_{\rm jet}}}\right)^2
+ \left(\frac{M_{\rm jet}^{\rm recoil}-m_{Z^0}}
  {\sigma_{M_{\rm jet}^{\rm recoil}}}\right)^2\]
is less than 4.5, where $M_{\rm jet}$ and $M_{\rm jet}^{\rm recoil}$ 
are the invariant mass and the
recoil mass reconstructed from the two jets.
After selecting the $Z^0H$ candidates, cuts of    
$N_{track} \geq$ 10 and $Y_{cut}^{4jet} \leq $ 0.012
are applied to reduce the $H\to\tau^+\tau^-,W^+W^-$ backgrounds,
respectively,
similar to the case of the 4 jet analysis.
The expected numbers of events in a year and the selection 
efficiencies are summarized in Table~\ref{seleZH}.

%
%
\subsection{2 Jet + {\it ll} Analysis}\label{sec:2ljet}
\begin{figure}[t]
\epsfxsize=.48\linewidth
\centerline{\epsfbox{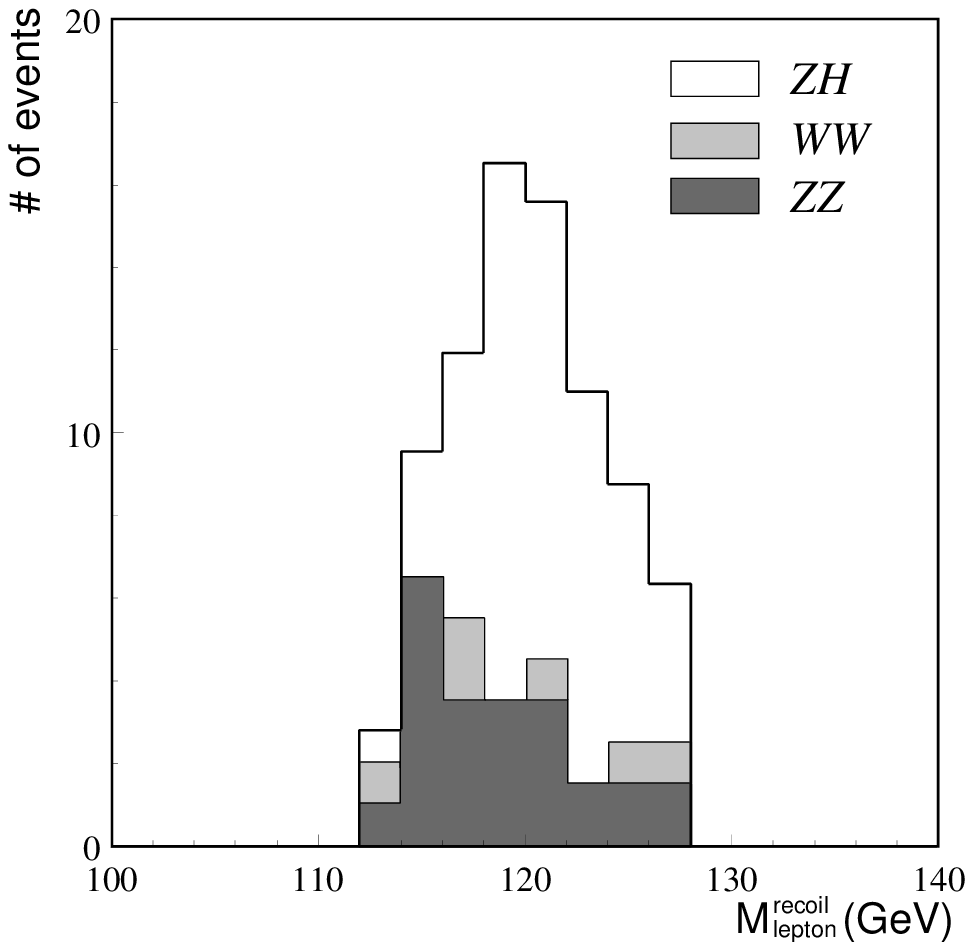}}
\hangcaption{$M_{\rm lepton}^{\rm recoil}$ distribution after all cuts for
	2 jet+{\it ll} analysis. The shaded histograms are for the
	background events.} 
\label{2ljetZH}
\end{figure}

In this analysis, two same flavor leptons ($\rm e,\mu$) with opposite
signs are required in an event. 
After removing the two lepton tracks, all events are reconstructed
as a 2 jet final state.
Containment cuts of $|\cos{\theta_{\rm jet}}| \leq $0.8 and 
$|\cos{\theta_{\rm lepton}}| \leq $0.8 are applied. 
Events which satisfy the condition 
\[\Delta_{ZH} = 
  \left(\frac{M_{\rm lepton}^{\rm recoil}-m_{H}}
	{\sigma_{M_{\rm lepton}^{\rm recoil}}}\right)^2
+ \left(\frac{M_{\rm lepton}-m_{Z^0}}
	{\sigma_{M_{\rm lepton}}}\right)^2 \leq 7 \]
are selected as $Z^0H$ candidates,
where $M_{\rm lepton}$ and $M_{\rm lepton}^{\rm recoil}$
are the invariant mass and the recoil mass
reconstructed from the lepton pair.
We use here $M_{\rm lepton}^{\rm recoil}$ as the Higgs mass.
The same cuts as the 2 jet analysis are applied
to reduce the background events from $H\to\tau^+\tau^-,W^+W^-$.
The distribution of $M_{\rm lepton}^{\rm recoil}$ in this analysis  
is shown in Fig.~\ref{2ljetZH}.

The selection criteria for the $\sigma_{tot}$ measurement
is almost the same as 
that of 2jet+{\it ll} analysis except that the cut for
$\cos{\theta_{\rm jet}}$ is not used.  
The expected numbers of events in a year and the selection 
efficiencies are summarized in Table~\ref{seleZH}.

\begin{table}[H]
\caption{The event selection efficiencies and expected 
	numbers of the $Z^0H$ events in a year (50 fb$^{-1}$) for the 
	SM Higgs boson with a mass of 120 GeV.
	Those of the background processes are also given.}
\label{seleZH}	
\begin{center}
\begin{tabular}{l*{8}{r}}\hline\hline
 & \multicolumn{2}{c}{4 jet} & \multicolumn{2}{c}{2 jet} 
& \multicolumn{2}{c}{2 jet+{\it ll}} 
& \multicolumn{2}{c}{$\sigma_{tot}$} \\
 & $\varepsilon_{\rm sel}$ (\%) &\# events 
& $\varepsilon_{\rm sel}$ (\%)&\# events 
& $\varepsilon_{\rm sel}$ (\%) & \# events 
& $\varepsilon_{\rm sel}$ (\%) & \# events \\ 
\hline
$e^+e^-\to Z^0H$  & 8.9&829&2.4&228 & 0.5 & 51.4 & 1.0& 98.7 \\ 
~~~~$H\to b\bar b$ &10.4&655&2.7&172 & 0.7 & 42.4 & 1.1& 67.8 \\ 
~~~~$H\to c\bar c$ &13.2& 30.3&4.4& 10.1 & 0.8 &  1.8 & 1.3& 3.0 \\ 
~~~~$H\to gg$      &12.8&102&4.5& 35.4 & 0.5 &  4.1 & 1.1& 8.6 \\ 
~~~~$H\to c\bar c+gg$ &13.0&132&4.4& 45.7 & 0.6 &  5.9 & 1.1& 11.6 \\ 
~~~~$H\to\tau^+\tau^-$ & 0.1&  1.0&0.0&  0.1 & 0.0 &  0.0 & 1.0& 6.9 \\ 
~~~~$H\to W^+W^-$ & 3.2& 41.2&0.8&10.6 & 0.2 &  3.1 & 1.0& 12.4\\
\hline 
$e^+e^-\to W^+W^-$  & 0.1&415&0.0& 43.0 & 0.0 &  6.0 & 0.0& 27.0\\ 
$e^+e^-\to Z^0Z^0$  & 0.4&233&0.1& 32.0 & 0.2 & 22.5 & 0.1& 70.5\\
$e^+e^-\to e^+e^-Z^0$& 0.0&  0.0&0.0&  0.0 & 0.0 &  0.0 & 0.0& 0.0\\
$e^+e^-\to e\nu W$  & 0.0&  0.0&0.0&  0.0 & 0.0 &  0.0 & 0.0& 0.0 \\ \hline\hline
\end{tabular}
\end{center}
\end{table}
%
%
\section{Flavor Tagging}
The technique of the flavor tagging used here
is based on the long lifetimes 
of hadrons containing heavy flavor quarks. 
Tracks from the decay of such hadrons have 
large impact parameters ($b$) from the primary vertex. 
The impact parameter 
can be precisely measured with the VTX surrounding the beam pipe.

\begin{figure}[H]
\epsfxsize=.5\linewidth
\centerline{\epsfbox{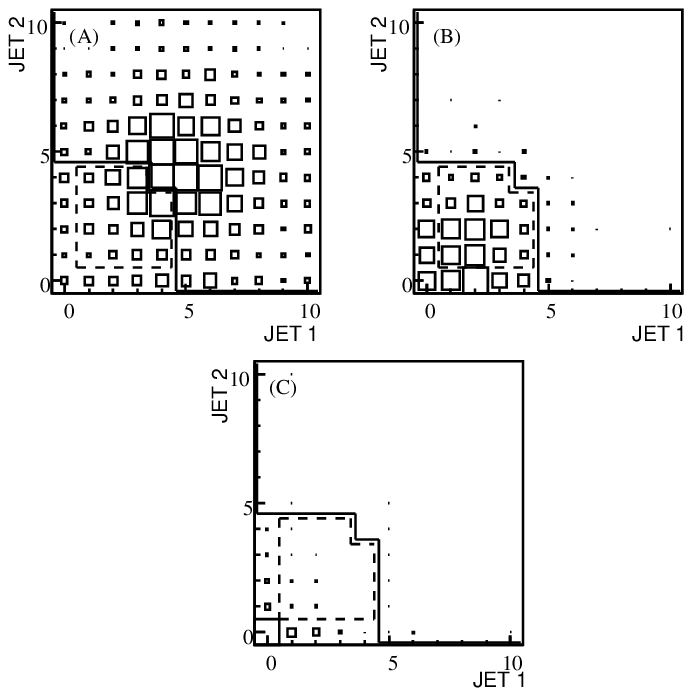}}
\hangcaption[]{Numbers of large impact parameter tracks in jets from 
	the Higgs decay for three Higgs decay modes 
	(A): $H\to b\bar b$, (B): $H\to c\bar c$ and 
	(C): $H\to gg$. The  outside events of solid line are
	assigned to $b\bar b$, and inside ones are assigned to
	$c\bar c, gg$. Inside of the dashed line is assigned to
	$c\bar c$.} 
\label{btagfig}
\end{figure}

In this study, we calculate the three dimensional impact parameter
using the space point information from the VTX and 
the momentum information from the CDC.
The error of the impact parameter measurement ($\sigma_b$) is
calculated according to the following formula~\cite{JLC-I}.
	\[\sigma_b = \sqrt{11.4^2 + 
	\left(\frac{28.8}{p\sqrt{\sin^3\theta}}\right)^3} (\mu m)
	~~~~~~( p~{\rm in~GeV)}\]
The first term in the square root is the geometrical contribution 
which is determined by the pixel size and the geometry  
of the VTX, while
the second term is the contribution of multiple scattering
which depends on the thickness of the detector.
We use the normalized impact parameter $b/\sigma_b$ to 
scale the impact parameter, instead of the impact parameter itself.
The number of tracks in the Higgs decay,
with their normalized impact parameters being 
greater than $\sigma_{cut}$, is used for the flavor tagging.
Fig.~\ref{btagfig} shows the distribution of 
the number of such tracks for the three Higgs decay modes 
together with the selection regions.
The value $\sigma_{cut}$ and the selection regions are optimized to
minimize the statistical error of 
$Br(H\to c\bar c, gg)/Br(H\to b\bar b)$.  
We use $\sigma_{cut}=3$ here.
The tagging efficiency\footnote{%
Tagging efficiency is defined as:
$\varepsilon_{\rm tag} = \frac{\displaystyle{ \rm \#~of~tagged~events}}
{\displaystyle{{\rm \#~of}~Z^0H~{\rm events}}}$}
and the expected number of tagged events in a year 
are summarized in Table~\ref{btagtab} for each final state.

\begin{table}[H]
\caption{The tagging efficiencies and expected numbers of signal and 
 	background events for the SM Higgs boson with a mass of 120
	GeV (50 fb$^{-1}$).  Here $\sigma_{cut} = 3$ 
	are used. Efficiencies are given in percent. The bold
	numbers are efficiencies of the signal events.}
\label{btagtab}
\begin{center}
\begin{tabular}{l*{6}{r}}\hline \hline
\multicolumn{1}{c}{\smash{\lower2.ex\hbox{ $m_{H}$=120 GeV}} }
& \multicolumn{3}{c}{b-tag (\%)}
&\multicolumn{3}{c}{$c\bar c, gg$-tag (\%)} \\ 
& 4 jet & 2 jet & 2 jet+{\it ll}& 4 jet & 2 jet & 2 jet+{\it ll} \\
\hline
$e^+e^- \to Z^0H$      &  &  \\ 
~~~~$H\to b\bar b$ &\bf 76.0 &\bf 85.5 &\bf 74.7& 24.0 & 14.5 & 25.3 \\ 
~~~~$H\to c\bar c$ &  9.4 & 10.0 &  9.0 & 90.6 & 90.0 & 91.0 \\ 
~~~~$H\to gg$       &  6.1 &  6.0 &  5.8 & 93.9 & 94.0 & 94.2 \\ 
~~~~$H\to c\bar c, gg$ &7.8 & 8.0 & 7.7 &\bf 92.2 &\bf 92.0 &\bf 92.3 \\ 
~~~~$H\to \tau^+\tau^-$ & 14.3 &  0.0 &  0.0 & 85.7 &100.0 &  0.0 \\ 
~~~~$H\to W^+W^-$    &  9.3 & 10.8 &  4.2 & 90.7 & 89.2 & 95.8 \\ 
\hline
$e^+e^-\to W^+W^-$    &  1.9 &  7.0 &  0.0 & 98.1 & 93.0 &100.0 \\ 
$e^+e^-\to Z^0Z^0$   & 14.8 & 18.7 & 20.0 & 80.4 & 81.3 & 80.0 \\ 
\hline
Signal/B.G.             &497/67&147/13& 32/5 &123/789&43/101&6/38 \\ \hline \hline
\end{tabular}
\end{center}
\end{table}
%
%
\section{Selection of $H\to\tau^+\tau^-$}

The event of the process $e^+e^-\to Z^0H$ 
with the decay $H\to\tau^+\tau^-$
has at least two undetectable neutrinos in the final state. 
Therefore the $Z^0H$ event of this decay mode should be identified 
using mainly $Z^0$ decay product, and the 2 jet events where $Z^0$
decays into two neutrinos are not used in this analysis.
The strategy of the event selection is the following:
\begin{itemize}
\item Select $Z^0H$ event using the $Z^0$ decay product and 
	the opening angle ($\theta_H$) of the jets from the Higgs
	decay. 
\item Select events with a decay $H\to\tau^+\tau^-$
	by using the missing energy and the charged multiplicity. 
\end{itemize}

In the 4 jet analysis, we select a pair of jets
whose invariant mass is closest to $m_{Z^0}$. 
Cuts of  $|\cos{\theta_{\rm thrust}}| \leq$ 0.8, $Thrust
\leq$ 0.8, $E_{\rm mis} \geq$ 10 GeV, and
$\theta_H \geq 90^\circ$ are then applied. 
We require that the recoil mass against the $Z^0$ 
is consistent with the Higgs mass, and that each jet from the Higgs
decay contains at most five charged tracks.

In the 2 jet+{\it ll} analysis, 
we require that there is a pair of opposite sign leptons in the event,
and that the invariant mass and the recoil mass of the leptons
are consistent with $Z^0$ and $H$, respectively.
Cut on the missing energy of $E_{\rm mis} \geq$ 10 GeV is then applied.
The two jets are considered to be the Higgs decay product
and the number of charged tracks in each jet
should be less than three. The selection efficiencies and 
the expected numbers of signal and background
events in a year are summarized in Table~\ref{tautab}.
\pagebreak
\begin{table}[H]
\caption{The selection efficiencies and expected 
	numbers of the $H\to\tau^+\tau^-$ events for the SM Higgs
 	boson with a mass of 120 GeV (50 fb$^{-1}$).
	Those of background events are also given.}
\begin{center}
\begin{tabular}{l*{7}{r}} \hline\hline
 &\multicolumn{2}{c}{4 jet} &\multicolumn{2}{c}{2 jet} 
&\multicolumn{2}{c}{2 jet+{\it ll}} \\
 & $\varepsilon_{\rm sel}$ (\%) & \# events 
& $\varepsilon_{\rm sel}$ (\%) & \# events
& $\varepsilon_{\rm sel}$ (\%) & \# events \\ \hline
$e^+e^- \to Z^0H$    & \\ 
~~~~$H\to b\bar b$  &$\sim 0$&  0.4 & --- & --- &$\sim 0$& 0.0 \\ 
~~~~$H\to c\bar c$  &$\sim 0$&  0.2 & --- & --- &$\sim 0$& 0.0 \\ 
~~~~$H\to gg$       &$\sim 0$&  0.0 & --- & --- &$\sim 0$& 0.0 \\ 
~~~~$H\to c\bar c, gg$ &$\sim 0$&  0.2 & --- & --- &$\sim 0$& 0.0 \\ 
~~~~$H\to\tau^+\tau^-$&{\bf 11.8}&{\bf 74.9}&---&---&{\bf 0.8}&{\bf 4.9}\\ 
~~~~$H\to W^+W^-$  &   0.3  &  3.1 & --- & --- &    0.1 & 1.1 \\ \hline
$e^+e^-\to W^+W^-$  &$\sim 0$&  1.0 & --- & --- &$\sim 0$&10.0 \\ 
$e^+e^-\to Z^0Z^0$   &$\sim 0$&  9.5 & --- & --- &$\sim 0$& 0.0 \\
$e^+e^-\to e^+e^-Z^0$ &$\sim 0$&  0.0 & --- & --- &$\sim 0$& 0.0\\
$e^+e^-\to e\nu W$   &$\sim 0$&  0.0 & --- & --- &$\sim 0$& 0.0\\\hline
Signal/B.G.           &   & 74.9/14.4 & --- & --- &   & 4.9/11.1 \\\hline \hline
\end{tabular}
\label{tautab}
\end{center}
\end{table}
%
%
\section{Results}
\begin{figure}[H]
\epsfxsize=.7\linewidth
\centerline{\epsfbox{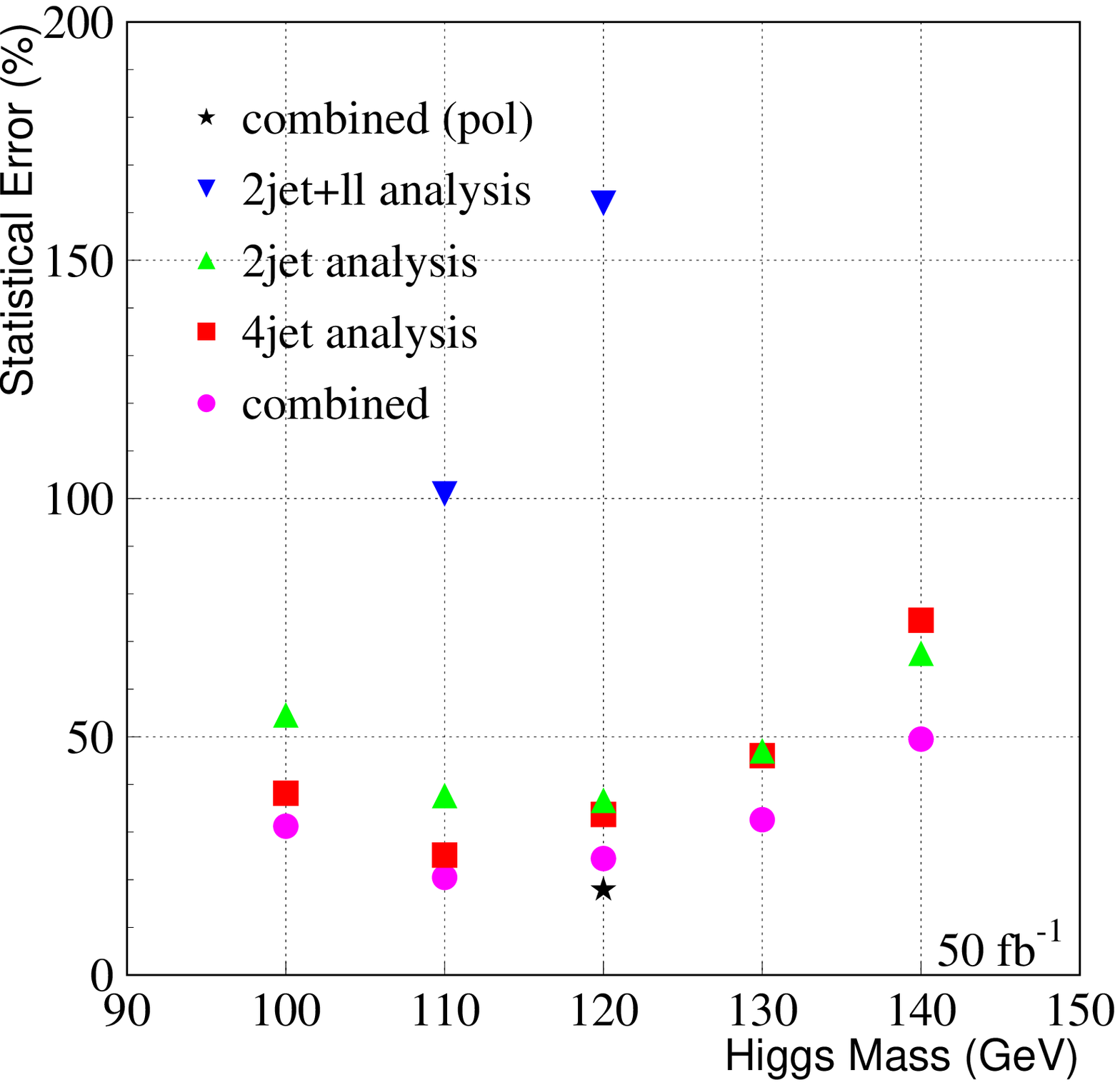}}
\hangcaption[]{Statistical errors of 
	$Br(H\to c\bar c, gg)/Br(H\to b\bar b)$ as a function of 
	the Higgs mass.}
\label{edep}
\end{figure}

We list in Tables~\ref{seleZH}, \ref{btagtab} and \ref{tautab}
the expected numbers of signal and background events 
to be collected in a year. 
Using these numbers, statistical errors 
of the total cross section 
times branching fractions
and the ratios of the branching fractions are calculated.
The results are listed in Table~\ref{res-sum}.
As shown in this table, the results of the 2 jet analysis are 
comparable to those of the 4 jet analysis. 
The combined results show significant improvement
compared to the results shown in Ref.~\cite{hildreth}. 
It is found that the beam polarization gives rather modest improvement
to the measurement of $Br(H\to c\bar c, gg)/Br(H\to b\bar b)$. 
Similar studies are performed also for different Higgs masses.
The statistical errors of $Br(H\to c\bar c, gg)/Br(H\to b\bar b)$ 
are shown in Fig.~\ref{edep} as a function of the Higgs mass.
At the low Higgs mass region, irreducible background from 
$e^+e^-\to Z^0Z^0$ increases, while the production cross section
decreases in the high Higgs mass region. Thus the best result 
will be obtained if the Higgs mass lies around 110-120 GeV. 

\begin{figure}[H]
\epsfxsize=.7\linewidth
\centerline{\epsfbox{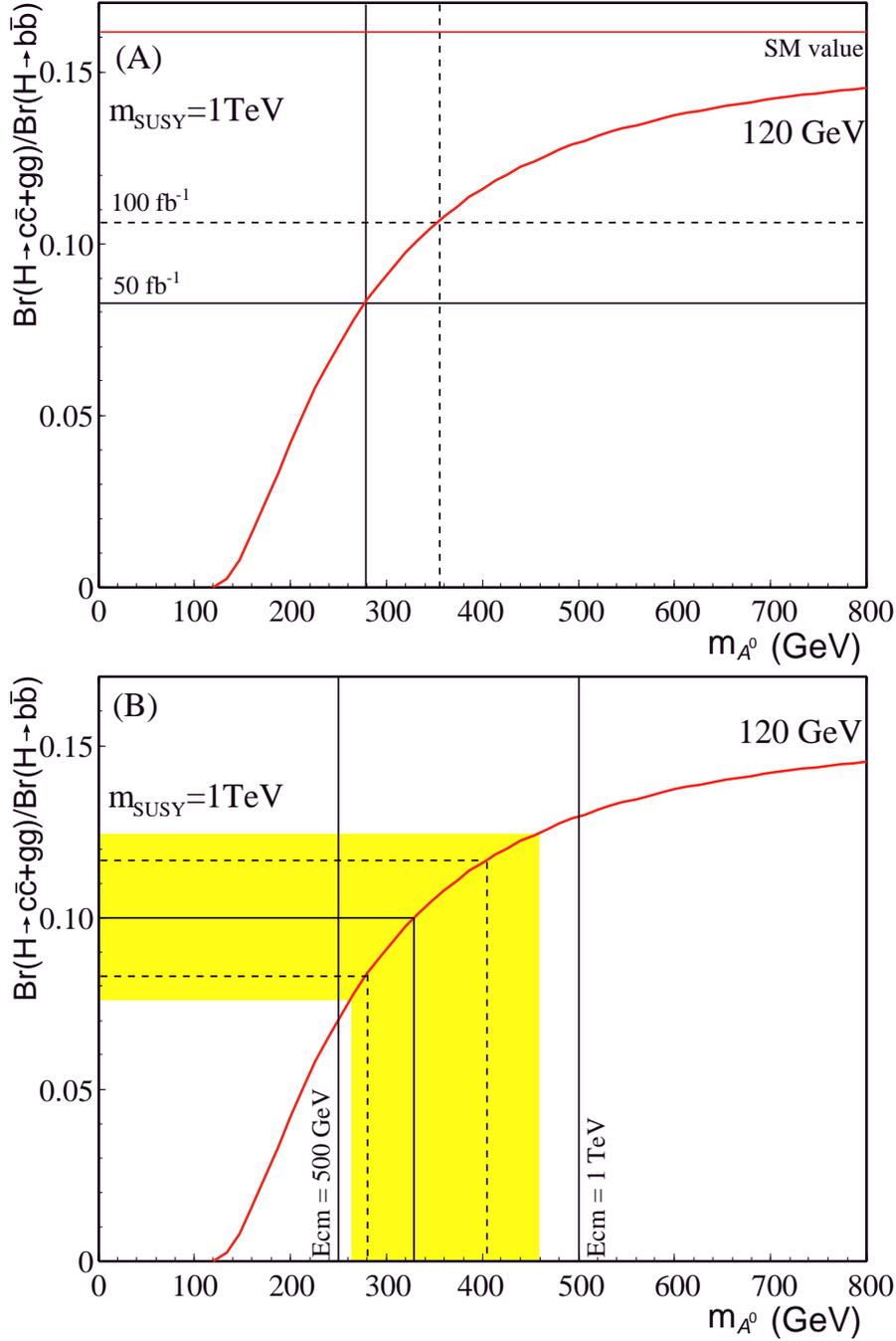}}
\caption[]{Relation between $Br(H\to c\bar c, gg)/Br(H\to b\bar b)$
	 and $m_{A^0}$ for the Higgs boson with a mass of 120 GeV.
	(A) The solid (dotted) lines show 95 \% confidence level lower
	limit after 1 year (2 years). 
	(B) The shaded area is the allowed region after a year 
	(1 standard deviation). The region in the dotted lines 
	shows the allowed region after two years. 
	Kinematical thresholds for $A^0$ direct production
	are also drawn.}
\label{branch-azero2}
\end{figure}

Fig.~\ref{branch-azero2}A shows the relation between 
$Br(H\to c\bar c, gg)/Br(H\to b\bar b)$ and $m_{A^0}$
together with 95 \% confidence level lower limit 
for the SM Higgs boson with a mass of 120 GeV.
If the ratio were measured to be 0.162 $\pm$ 0.028
with an integrated luminosity of 100 fb$^{-1}$
(the SM prediction value with a relative error of 17 \%),
the MSSM which gives the ratio less than 0.106
will be excluded at 95 \% confidence level.
In other words, a lower limit on $m_{A^0}$
will be set to be 360 GeV at 95 \% confidence level.
To the contrary, if the ratio were measured to be significantly
lower than the SM value,
the SM will be excluded and
both lower and upper limits on $m_{A^0}$ will be obtained.
Fig.~\ref{branch-azero2}B shows such a case where
the ratio were measured to be 0.100
with a similar relative accuracy.

\begin{figure}
\epsfxsize=.6\linewidth
\centerline{\epsfbox{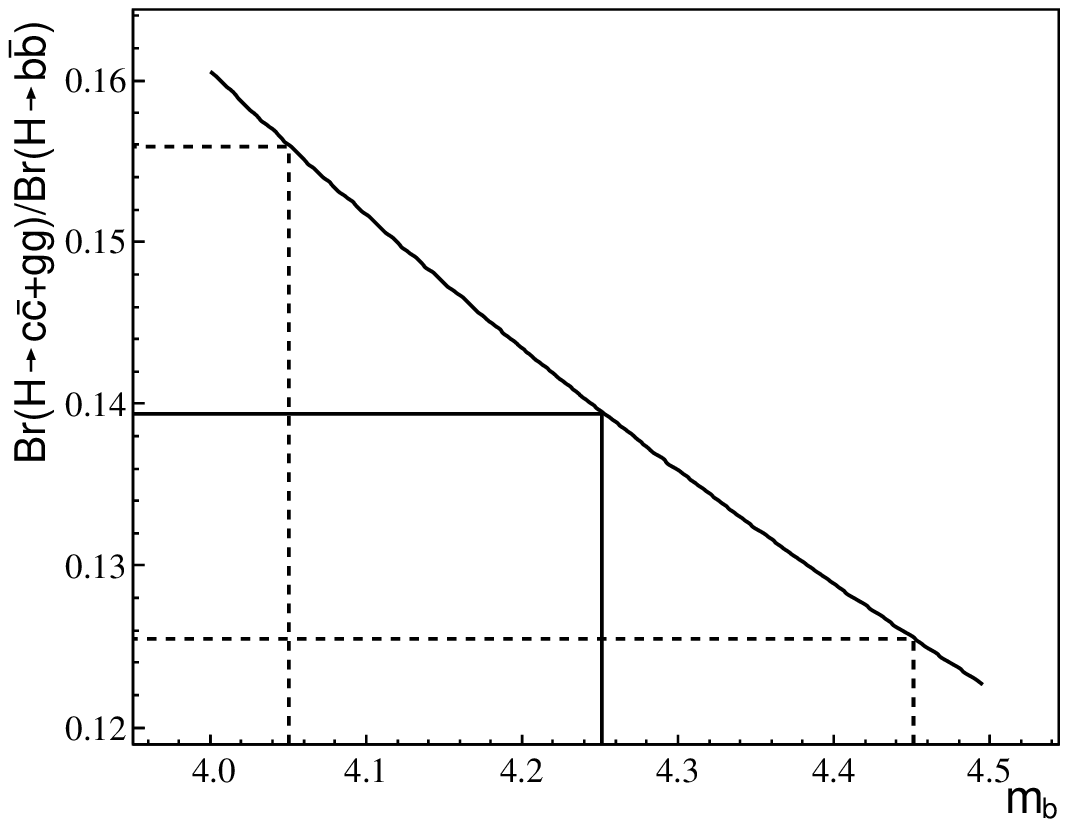}}
\hangcaption[]{The relation between $Br(H\to c\bar c, gg)/Br(H\to b\bar b)$
	 and $m_b$. The dotted lines	show the uncertainty after two
	years.} 
\label{taufig}
\end{figure}

\begin{table}[H]
\caption{Expected measurement errors of various values after a year
	(50 fb$^{-1}$) for the SM Higgs boson with a mass of 120 GeV (\%). }
\label{res-sum}
\begin{center}
\begin{tabular}{c*{8}{r}}\hline\hline
&\multicolumn{4}{c}{no polarization}
&\multicolumn{4}{c}{90 \% polarization} \\
$\frac{\Delta X}{X}$ & 4 jet & 2 jet & 2 jet+{\it ll} 
& combined & 4 jet & 2 jet & 2 jet+{\it ll} & combined \\ 
\hline
$\sigma_{tot}$ &---&---&24.2&---&---&---&22.7&--- \\ 
$\sigma_{tot}\times Br(H\to b\bar b)$ 
& 5.2 & 9.0 & 20.9 & 4.4 & 5.2 & 7.3 & 18.6 & 4.1 \\
$ \sigma_{tot}\times Br(H\to c\bar c)$ 
& 145 & 161 & 532 & 105 & 107 & 127 & 338 & 79.5 \\
$\sigma_{tot}\times Br(H\to c\bar c+gg)$ 
& 33.3 & 35.6 & 161 & 24.0 &  24.1 & 26.0 & 96.5 & 17.4\\
$ \sigma_{tot}\times Br(H\to\tau^+\tau^-)$
& 13.6 & ---  & 87.2 & 13.4 & 14.9 & --- & 95.7& 14.7 \\
$\frac{\displaystyle Br(H\to c\bar c)}{\displaystyle Br(H\to b\bar b)}$ 
& 70.1 & 77.1 & 159 & 49.3 & 56.0  & 68.6 & 250 & 42.7 \\
$\frac{\displaystyle Br(H\to c\bar c+gg)}
{\displaystyle Br(H\to b\bar b)}$ 
& 33.7 & 36.7 & 162 &  24.5 & 24.6 & 27.0 & 98.2 & 17.9 \\ \hline\hline
\end{tabular}
\end{center}
\end{table}
\pagebreak
\vspace*{1cm}
\pagebreak
The fact that $Br(H\to\tau^+\tau^-)$ can be measured with a
statistical error of 13.4 \% corresponds to that $m_b$ is determined
with an error of 6.7 \%. The systematic error of 
$Br(H\to c\bar c, gg)/Br(H\to b\bar b)$ from the uncertainty of $m_b$
is estimated to be $\sim$ 10 \% after two years. Fig~\ref{taufig} shows
the relation between $Br(H\to c\bar c, gg)/Br(H\to b\bar b)$ and $m_b$.

%
%
\section{Summary}
We performed a detailed simulation study for the measurement of the
ratio $Br(H\to c\bar c, gg)/Br(H\to b\bar b)$
at $\sqrt{s}=300$ GeV assuming the JLC standard detector. 
Here we introduce new approaches to use
all decay modes of $Z^0$ and to consider the beam polarization.
We find that the results obtained from the 2 jet analysis are
comparable to those of the 4 jet analysis,
although the original event ratio is 2~jet : 4~jet $\sim$ 20 : 70.
Thus we can reduce the statistical error by combining
the results of all decay modes.
On the other hand, the effect of the beam polarization is 
found to be small for this analysis.

The statistical error of the ratio $Br(H\to c\bar c, gg)/Br(H\to b\bar b)$
after two years is estimated to be 17 \% for the SM Higgs
with a mass of 120 GeV. 
With the good accuracy of the measurement,
we have a chance
to rule out the SM, 
or to constrain the mass of $A^0$ in the MSSM,
according to the measured value of the ratio.
%
%
\section*{acknowledgments}
We would like to thank Prof. Yasuhiro Okada, Drs. Junichi Kamoshita, Minoru
Tanaka and other members of the JLC Higgs Working Group for many
useful discussions. We would like to also thank Drs. Junichi Kanzaki
and Keisuke Fujii for great help for development of the detector simulator. 
%
%

\end{document}